# Nanoparticle Photo-Ejection from Liquid via Excited Plasmonic Supercavitation

Qiushi Zhang[1], Dezhao Huang[1], Seunghyun Moon[1], Jarrod Schiffbauer[2], Eungkyu Lee[3,*], and Tengfei Luo[1,4,*]

1. Department of Aerospace and Mechanical Engineering, University of Notre Dame, IN, USA

2. Department of Physical and Environmental Sciences, Colorado Mesa University, Co, USA

3. Department of Electronic Engineering, Kyung Hee University, Yongin-si, South Korea

4. Department of Chemical and Biomolecular Engineering, University of Notre Dame, IN, USA

* Corresponding authors: eleest@khu.ac.kr; tluo@nd.edu.

**Abstract:**

The ability to separate miniscule solid particles (*e.g.*, nanoparticles, NPs) from liquid is important to a wide range of applications, such as water purification, material deposition, and biomedical engineering. Such separation is usually achieved by displacing liquid via filtration or distillation. However, directly moving small particles out of liquid is difficult, especially when their sizes approach the nanometer scale, as the capillary force on the NP at the liquid surface is too large for common body forces (*e.g.*, optical or magnetic) to overcome. Here, we demonstrate the ability to eject metallic NPs out of liquid with a laser excitation at their surface plasmon resonance (SPR) wavelength. The laser applies an optical force on the NPs to drive them toward the liquid surface. In the meantime, the laser can also intensely heat the NP to form a nanobubble encapsulating the NP (*i.e.*, supercavitation), which achieves the liquid-NP separation and thus eliminates the capillary force on the NP at the liquid free surface. We show that such a mechanism can expel NPs out of liquid as observed using a transient scattering experiment, which is further confirmed by molecular dynamics simulations. We also demonstrate depositing the NPs on a solid surface not in contact with the liquid. This study reveals an interesting mechanism to separate NPs from liquid and could potentially benefit separation, nanomaterials and biomedical applications.

**Introduction**

The ability to separate miniscule solid particles from liquid is essential to a wide range of applications that need particle seperation,[1–5] concentration[6–11] and deposition.[12–21] For example, to study the toxicological effects of engineered nanoparticle (NP) aerosols on the human respiratory system,[22–25] direct deposition of solid NPs onto human cells would be ideal for in-vitro experiments, but current technologies[26,27] mostly rely on using aerosols to carry such NPs.[28,29] Existing methods for separating solid particles from the hosting liquids, such as filtration and distillation, usually achieve solid-liquid separation by displacing liquids, *i.e.*, passing liquids through a membrane or evaporating it into vapor. However, directly moving small particles out of liquid is more challenging, especially when their sizes approach the nanometer scale. These tiny particles can be stranded at the liquid/air interface because of the capillary force, which has led to applications such as self-assembly.[30–33] But such forces are so strong, scaling with the inverse of particle radius, that common body forces are too weak to drive small particles out of the liquid. For example, for a NP with a diameter of 120 nm, the capillary force[34,35] on it would be ~$10^{-8}$ N at the water/air interface (see Supporting Information, SI1, for calculation details). However, body forces like optical scattering forces and magnetic forces commonly used to drive suspended NPs are many orders of magnitude smaller than the capillary force. For example, the dispersive optical scattering force on a 120-nm-diameter gold/silica core shell NP is ~$10^{-12}$ N even with a relatively high optical fluence of 9~15 mJ/cm$^2$,[36–39] and the magnetic force on a 10-nm-diameter colloidal iron oxide NP in magnetic fields with strengths of 5~15 Tesla is only ~$10^{-18}$ N.[40] However, if there is a way to separate surrounding liquid from the solid NP surface prior to reaching the liquid interface, there will be a chance that the NP can escape from the liquid without being stranded by the capillary force. It has been demonstrated that optically excited localized surface plasmonic heating can lead to nanoscale vapor bubbles to encapsulate the NPs, *i.e.*, supercavitating NPs, if the optical fluence is above a certain threshold,[41–45] which may help realize the desired liquid-NP separation. These supercavitating NPs can also be driven by optical scattering forces toward the liquid interface guided by the light.[36,38,46]

In this work, we demonstrate the ability to eject metallic NPs out of liquid with a laser excitation at the surface plasmon resonance (SPR) wavelength of the NPs. The laser is shown to function as both a supercavitation exciter and an optical force provider, that creates the aforementioned liquid-NP separation and in the meantime drives the supercavitating NPs toward the liquid interface. We show that the optically driven supercavitating NPs can move out of liquid as observed using a transient scattering experiment, which is also verified by molecular dynamics (MD) simulations. We also demonstrate the NP expulsion by depositing them onto a glass surface not in contact with the liquid. Our temperature field analysis using finite element thermofluidic simulations confirm that the observed ejection of NPs from liquid does not originate from boiling or evaporation. Our study reveals a novel mechanism to enable NP-liquid separation and could potentially benefit separation, nanomaterials and biomedical applications.

**Results and discussion**

We first show experimentally that Au NPs in a suspension can be driven toward the air/liquid interface and expelled from liquid under the illumination of a laser at the SPR peak wavelength (see Methods section for experimental setup details). The optical system is shown in **Figure 1a**, in which a droplet of Au NP suspension with a concentration of ~$1\times10^{15}$ particles/m$^3$ is held by a thin glass substrate. The Au NP consists of a silica core (100 nm in diameter) and a thin Au shell (10 nm in thickness), supporting the SPR peak at ~800 nm in water. An 800 nm femtosecond pulsed laser is focused by a 20× objective lens onto the air/liquid interface at the tip of the droplet, which is the source laser used to excite the Au NPs to achieve supercavitation by leveraging the localized intense heating at the SPR. The side view of the droplet is monitored by a high-speed camera with a 10× objective lens. We use the dark-field scattering method[36,47] with an additional HeNe probe laser (2 mW, at the wavelength of 632.8 nm) illuminating the air side of the air/liquid interface around the focal point of the source laser to monitor the dynamics of the ejected NPs (see schematic in **Figure 1a**). We note that the intensity of the probe laser (0.64 W/cm$^2$) is very low and thus its weak optical force should not influence the dynamics of NPs. It is noted that we use an optical filter to block the source laser light to reach the image sensor of the camera in the dark-field scattering measurement. As shown in **Figure 1b** (left), the source laser is focused on the air/liquid interface, and the air region monitored by the camera field of view is illustrated by a dashed black square. In the right panel of **Figure 1b** (also see Supporting Movie, M1), where we used the source laser with an optical fluence of 22.8 mJ/cm$^2$ at the focal point, we can observe many red glowing dots in the dark field, which correspond to the diffraction-limited scattered light[36,38,46,47] of the probe laser from the Au NPs ejected into the air.

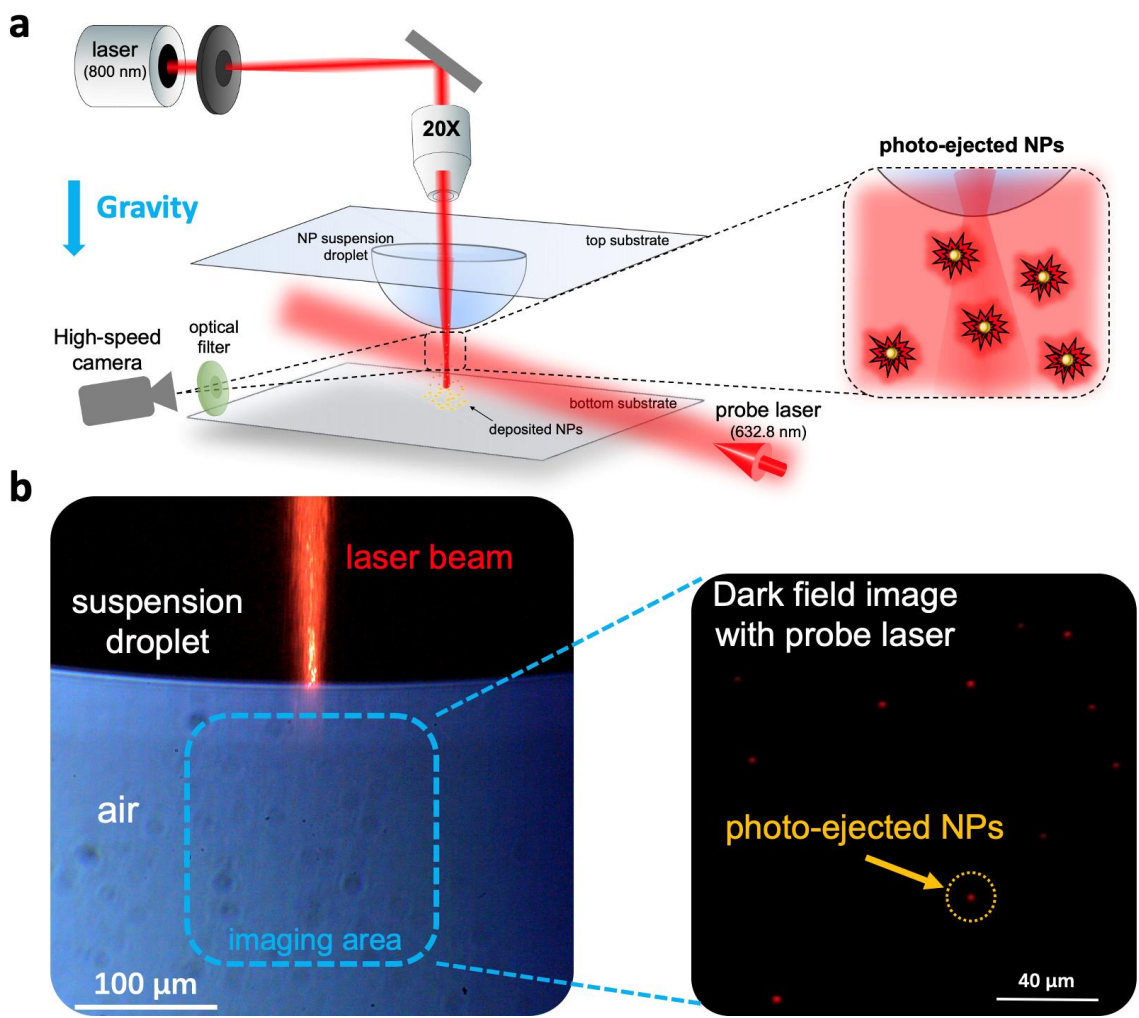

**Figure 1. Experimental system for monitoring the photo-ejected Au NPs across the air/liquid interface.** (a) Schematic of the experimental setup to observe the photo-ejected Au NPs into the air by a source laser at the SPR wavelength. Gravity is in the downwards direction. (b) Left: bright field optical image with an LED illumination backlight showing the suspension droplet and the source laser beam focused on the air/liquid interface. Right: dark field optical image of the air side of air/liquid interface, which captures the photo-ejected Au NPs. Each red

glowing dot corresponds to the diffraction-limited scattered light of the probe laser from a single Au NP.[36,46]

When the NP suspension is irradiated by the source laser, the dispersive optical scattering force can push the NPs toward the liquid interface.[36,46] This force originates from the back scattering of the incident photons on the NP surface, which gives the NP a momentum in the light propagation direction.[48–50] The amplitude of this force is around $1.1 \times 10^{-11}$ N with the highest optical fluence at the focal point of 22.8 mJ/cm$^2$ used in our experiment (see Supporting Information, SI2, for dispersive optical scattering force calculation details).[36] However, such a force is about three orders of magnitude smaller than the capillary force (~$10^{-8}$ N, see Supporting Information, SI1), which can strand the NP at the liquid/air interface (**Figure 2a**).

For the core-shell Au NP studied in this work, when the source laser has an optical fluence above a certain threshold (~7 mJ/cm$^2$),[44] a nanobubble can be generated surrounding the intensely heated plasmonic NP and encapsulate it in vapor (*i.e.*, supercavitation). Our femtosecond pulsed laser has a fluence of 22.8 mJ/cm$^2$ that is well above this threshold, and thus supercavitation can occur, as proven in our previous studies[36,46,47] as well as others.[43,44] While the optical dispersive force from the source laser can drive the NP toward the liquid/air interface, supercavitation separates the solid NP from the liquid via a thermally induced phase change process[45,51–53] before the NP reaches the liquid/air interface, which in turn eliminates the need to overcome the capillary force at the liquid/air interface (**Figure 2b**).

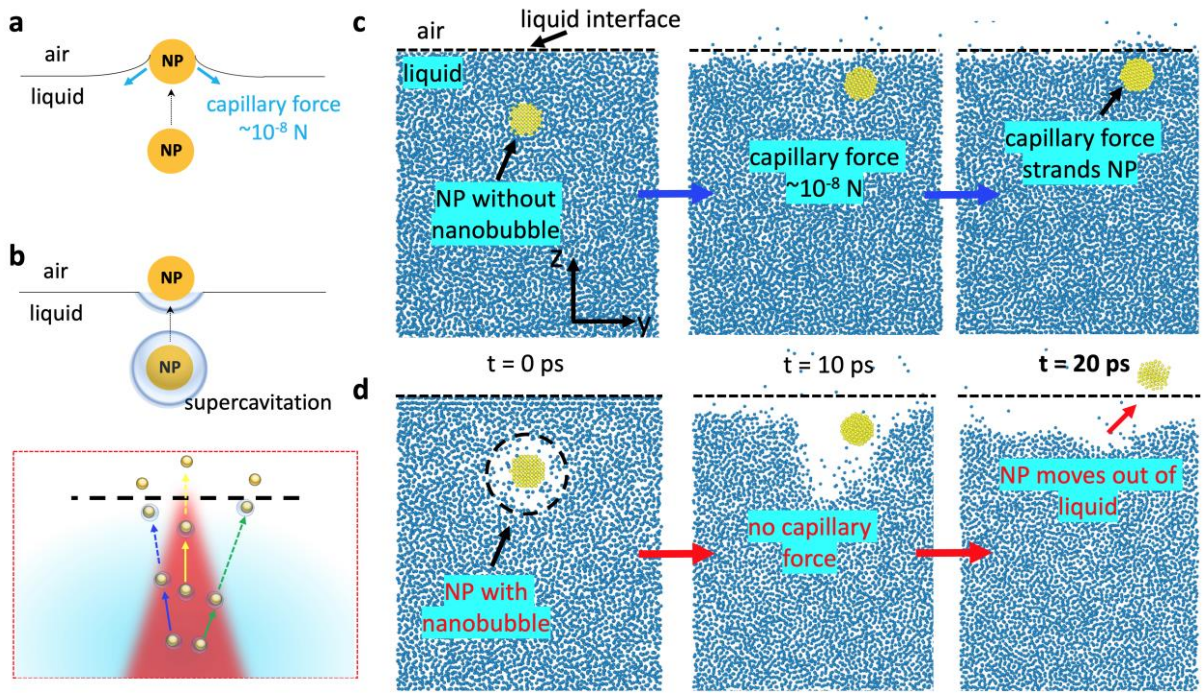

**Figure 2. Microscopic mechanism of supercavitating NP moving out of liquid interface.** Schematics showing (a) the NP stranded at the liquid/air interface due to the capillary force, but (b) with a supercavitation which separates NP from liquid within the liquid, the NP can pass through the interface without experiencing the capillary force at the liquid/air interface. Lower panel: The schematic of supercavitating NPs ejected out of liquid by laser. MD simulation snapshots of (c) a non-supercavitating NP moving toward and stranded at the interface, and (d) an intensely heated NP with supercavitation moving out of liquid.

To confirm this hypothesis, we perform a series of MD simulations of a solid NP immersed in liquid moving toward the liquid surface without and with thermally induced supercavitation (see Methods section for MD simulation details). We simulate an Au NP with a radius of 1 nm immersed in liquid argon, which has a free surface (**Figure 2c**). In one case, both the NP and the liquid are kept at 90 K, and thus no supercavitation is present. In the second case, the NP is heated to and maintained at 1000 K to excite a nanobubble encapsulating the NP.[54] We note that as long as the supercavitation can be generated, the exact heating temperature of the NP will not influence the conclusion of the simulations. As shown in **Figure 2c**, the NP is stranded by the liquid interface due to the capillary force (also Supporting Movie M2). However, when a supercavitation is generated to encapsulate the NP, the NP can move across the liquid interface without any impedance (**Figure 2d** and Supporting Movie M3). In this case, the NP-liquid separation is achieved when the nanobubble is generated, and when the NP approaches the liquid interface, there is no longer a capillary force holding back the NP from moving out of the liquid. We note that while these simulations are on a simplified model system of NP-in-liquid argon, the observation should be generally applicable to verify our hypothesis – it is the thermally excited supercavitation that enables the light to eject the NP from the liquid.

To further confirm that the observed laser-driven photo-ejection of Au NPs is due to the supercavitation, we need to exclude the effect of evaporation or boiling of the NP suspension due to the laser-induced volumetric heating,[39] which may also spread NPs into air from the suspension droplet (**Figure 1**). We perform thermofluidic simulations using the finite element method to calculate the steady-state temperature profile of a liquid droplet subject to laser heating due to the light absorption of the suspended plasmonic NPs. The simulation model is shown in **Figure 3a**, in which a water hemisphere with a radius of 1 mm is used to simulate the suspension droplet in the experiment. The plasmonic volumetric heating following a Gaussian distribution is the heating source in the system with the highest intensity located at the tip of the hemisphere, which mimics the focused laser beam (Gaussian beam) in our experiment.[38,39,46] The heating power of the system is determined by the laser optical fluence and NP concentration in the suspension. More details of the simulations can be found in Supporting Information, Section SI3. Since the highest temperature should occur at the laser focal spot, *i.e.*, the tip of the droplet, we plot the steady-state surface temperature profile of the hemisphere along the red line indicated in the model shown in **Figure 3a** to investigate the potential evaporation or boiling effect.

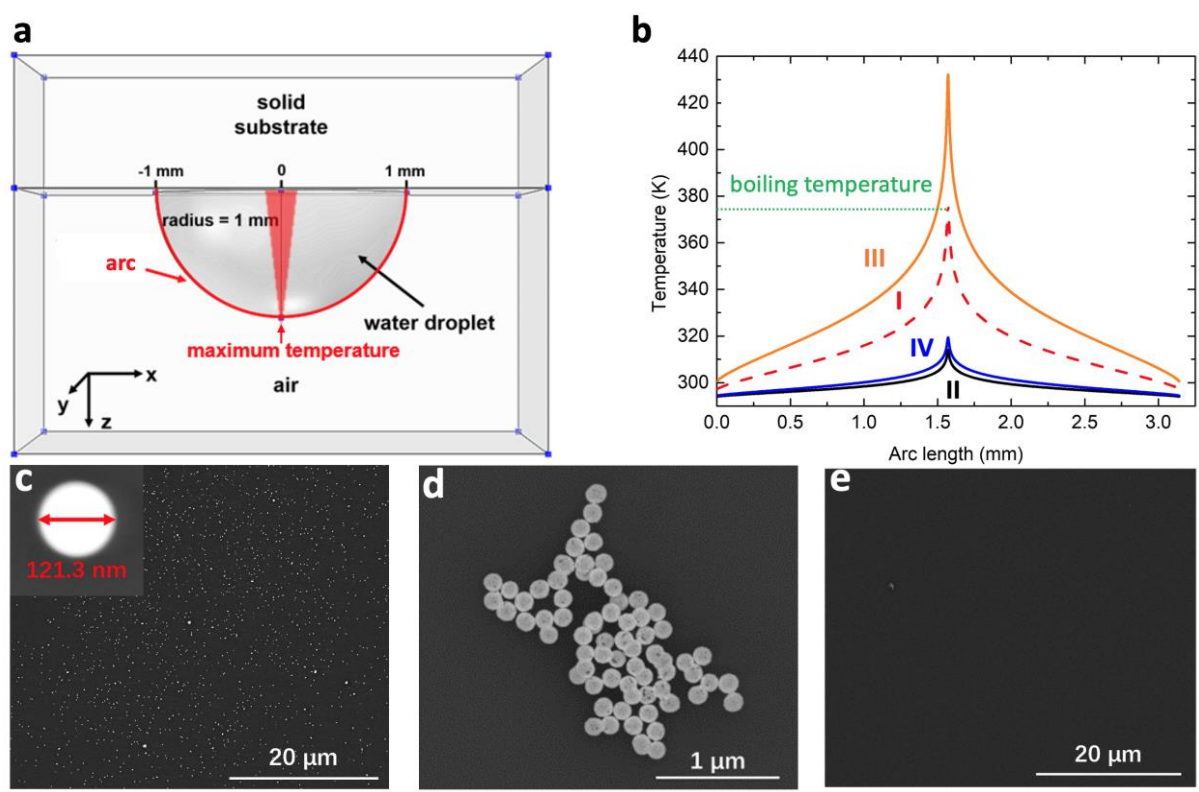

**Figure 3. Confirmation of supercavitation as the mechanism for the laser-driven photo-ejection of NPs out of liquid.** (a) The geometrical configuration of the model for thermofluidic simulations to calculate the surface temperature profile of an Au NP suspension droplet under the plasmonic volumetric heating effect. The region highlighted in red indicates the region

where the plasmonic volumetric heating occurs due to the illumination from a Gaussian beam. The red solid line depicts the arc line along which the surface temperature is visualized. (b) The steady-state surface temperature profiles of the droplet along the arc line in four cases with different optical fluences and NP concentrations to realize different plasmonic heating powers. Different cases are labeled by Roman numbers I-IV. I: 22.8 mJ/cm$^2$ and 2.2×10$^{15}$ particles/m$^3$, II: 22.8 mJ/cm$^2$ and 1.0×10$^{15}$ particles/m$^3$, III: 22.8 mJ/cm$^2$ and 3.0×10$^{15}$ particles/m$^3$, and IV: 5.4 mJ/cm$^2$ and 2.5×10$^{15}$ particles/m$^3$. The water boiling temperature of 373 K is indicated by a green dashed line. The back-scattered scanning electron microscope (SEM) images for cases (c) II, (d) III and (e) IV are shown. The insert in (c) highlights an individual Au NP (~120 nm diameter).

Four cases with different heating powers are studied by changing the optical fluence and NP concentrations. The temperature profiles of the four cases are plotted in **Figure 3b**. As expected, all the four cases have a symmetric surface temperature profile with the maxima located in the middle corresponding to the laser focal point. In case I (red dash line in **Figure 3b**), where the optical fluence and NP concentration are respectively 22.8 mJ/cm$^2$ and 2.2×10$^{15}$ particles/m$^3$, the maximum temperature is ~373 K, which is the threshold of water boiling. Case II (black solid line in **Figure 3b**) uses the same peak optical fluence (22.8 mJ/cm$^2$) but a lower NP concentration of 1.0×10$^{15}$ particles/m$^3$. These parameters are the same as we used in the experiment to visualize the NPs ejection in **Figure 1**. Because of the lower heating power in case II, the maximum temperature is only ~50 °C, *i.e.*, well below the boiling threshold, which means the observed Au NPs in air we previously imaged in **Figure 1b** (right) are unlikely caused by droplet boiling. To further investigate the Au NP photo-ejection mechanism, we placed a thin glass slide at a distance of 0.5 mm away from the tip of the suspension droplet (see **Figure 1a**), so that the NPs expelled out of the liquid by the laser can be deposited on the slide (see Methods section for experimental details) and then can be visualized using SEM. As the back-scattered SEM image shown in **Figure 3c** (also see the dark field optical images in Supporting Information, SI4), a large number of individual Au NPs are deposited on the glass slide. The insert in **Figure 3c** highlights a zoomed view of an individual NP, whose diameter is ~120 nm, and elemental composition (EDX result in Supporting Information, SI4) confirms that these are Au NPs from the suspension droplet.

As a comparison to case II, which has a maximum surface temperature below the boiling point, we study another case, case III, where the optical fluence is kept at 22.8 mJ/cm$^2$, but the NP concentration is increased to $3.0 \times 10^{15}$ particles/m$^3$. The resulted higher heating power leads to a maximum temperature of ~160 ℃, well above the boiling threshold (orange line in **Figure 3b**). In the experiment with the parameters corresponding to this case, we can also observe NPs deposited on the glass slide, but different from the scattered-distributed individual NPs in case II, we find many small clusters of NPs deposited on glass slide in case III. The back-scattered SEM image in **Figure 3d** shows an example of such clusters containing around ~100 NPs. As shown in our previous work,[36,46] supercavitating Au NPs moving in liquid do not tend to aggregate or form clusters at this concentration. Therefore, when they are ejected out of the liquid interface, they should be deposited individually as we see in case II (**Figure 3c**). However, when the heating power is sufficiently high to cause the droplet to boil as in case III, there can be tiny droplets splashing out of the suspension droplet. These tiny droplets can contain many NPs, and when they reach the glass slide and dry out, they can leave small clusters of NPs on the glass slide through a contact line deposition mechanism.[12,55]

While the maximum temperature in case II (~50 ℃) is not sufficient for boiling, the elevated temperature can increase evaporation. To ensure such enhanced evaporation cannot cause NPs to escape from the liquid droplet, we study another comparison case IV, where the peak optical fluence is reduced to 5.4 mJ/cm$^2$ but the NP concentration is increased to $2.5\times10^{15}$ particles/m$^3$, to still achieve the similar surface temperature as case II (blue solid line **Figure 3b**). We note that the peak optical fluence of 5.4 mJ/cm$^2$ is lower than the supercavitation threshold, 7 mJ/cm$^2$, as reported previously.[44] In this case IV, there is no NP deposited on the glass slide that can be observed in the back-scattered SEM (**Figure 3e**). This further confirms that supercavitation is the pre-requisite for the laser to expel the NPs out of liquid.

A potential application of this technique using photo-ejected NPs out of liquid is writing NP patterns on the surfaces that cannot be immersed in liquid solvents (*e.g.*, electronics). We use this photo-ejection enabled NP deposition technique to demonstrate the writing of a ~2 mm-long line of individual Au NPs on a glass substrate by translating the bottom substrate linearly while performing the deposition (see the schematic of setup in **Figure 4a** and SEM image in **Figure 4b**). The distance between the bottom glass substrate and the tip of the droplet is ~200 μm. Under this distance and NP concentration of $1.0\times10^{15}$ particles/m$^3$, the deposition rate is around a few hundreds of NPs per minute. It is interesting to see that the deposited NPs are spread out across the width of the written line with a range of ~330 μm as can be seen in **Figure 4b**. The density of deposited NP increases closer to the center of the written line. We believe the spread of NP is because the NPs are leaving the liquid interface over a range of angles and locations with respect to the laser beam axis. Using the dimensions shown in **Figure 4a**, we can estimate that the spread angle of the NP deposition is ~39.5°, which is similar to the estimated angle (35.8°) from the dark-field microscopy which tracks the locations of the NPs ejected out of the liquid (**Figure 4c**).

The $1/e^2$ diameter of the Gaussian source laser beam at the focal point, which is at the tip of the droplet, is ~12 μm, and the spread of the laser beam envelop after exiting the liquid interface (see Supporting Information, SI5, for calculation details) is much smaller than the observed spread of the trajectories of NPs coming out of the liquid (**Figure 4c**). Thus, the cause of the spread in the NP exit angle is unlikely to be caused by the divergence of the laser beam.

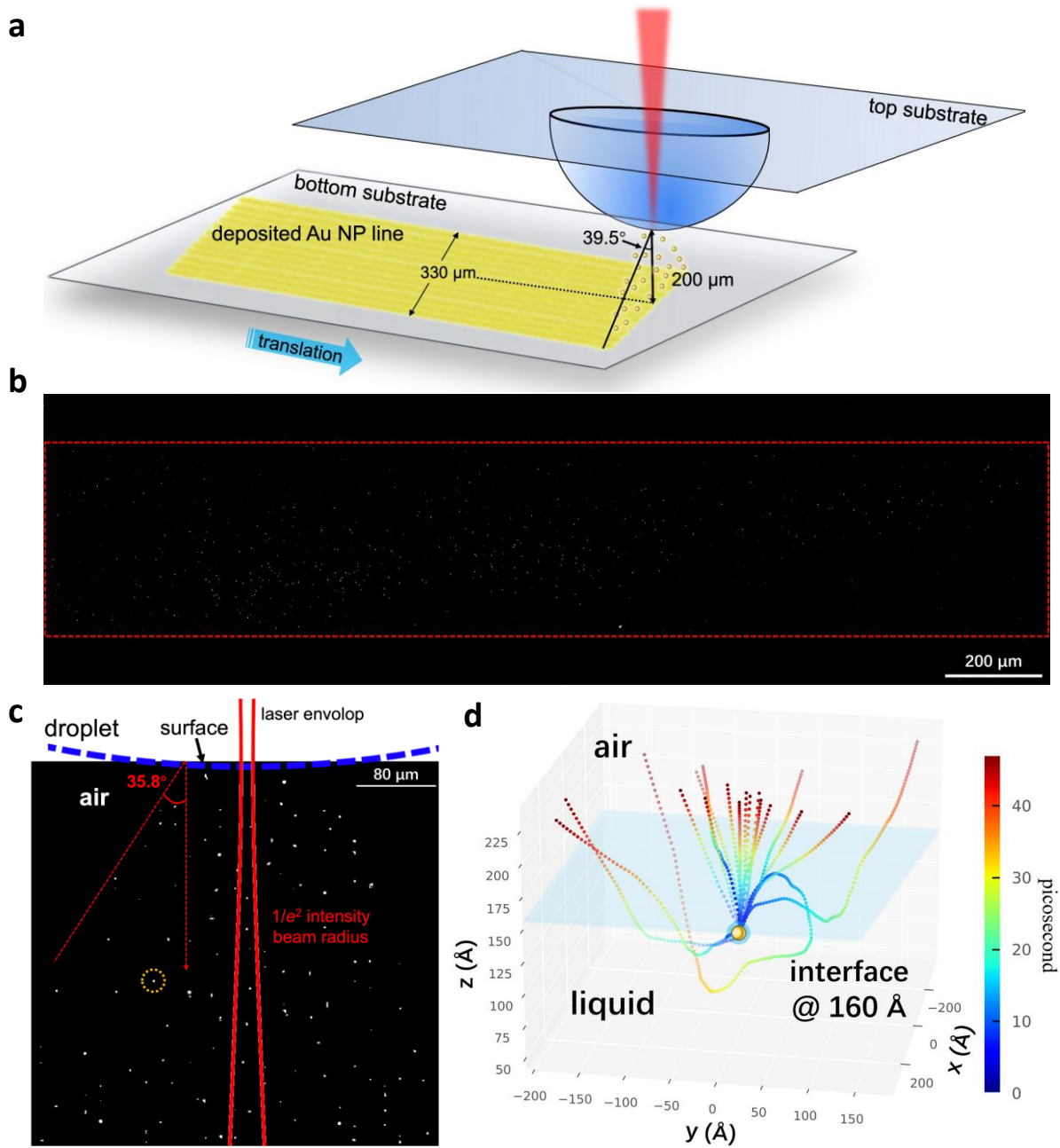

**Figure 4.** (a) Schematic of the NPs-line deposition experiment. (b) The back-scattered SEM image of the line of deposited Au NPs on the substrate by the laser with a fluence of 22.8 mJ/cm$^2$ and a NP concentration of $1.0\times10^{15}$ particles/m$^3$ in the droplet. The red dashed lines draw the approximated boundaries of the deposited line to guide the eyes. (c) The dark-field microscopy spatial distribution of the NPs photo-ejected into the air region (greyscale). The bright spots (circled in yellow) represent the locations of NPs ejected into air. The calculated

$1/e^2$ intensity profile of the Gaussian source laser beam is overlapped in the image. (d) The 3D MD-simulated trajectories of 18 different supercavitating NPs moving from liquid into air (interface is located at 160 Å).

We believe the spread angle is due to the stochastic nature of the nanobubble formation and the relative position of NP inside the nanobubble. It is known that the nanobubble formation is stochastic as the nucleation of the vapor bubble depends on the local temperature profile and the defects of the NP surface.[45,56] This can lead to the randomness in the fluidic forces on the NP, which can in turn change its moving direction.[54,57–59] Moreover, the relative position of NP inside the bubble is also stochastic, undergoing ballistic Brownian motion within the bubble,[54] and the randomness of the NP-bubble position can lead to different optical configurations and thus re-direct the dispersive optical scattering force direction on the NP.[36,47] As seen in Supporting Movie M4, supercavitating NPs can indeed move with some randomness in their directions, although in the long spatial range they still stay within the laser envelope, which is understandable as their movements are driven by the laser. When they approach the liquid interface, the fluctuations in the NP momentum causes deviation from the laser beam focusing axis as they come out of the liquid (see the schematic in **Figure 2b**). The same phenomena can also be observed in MD simulations where we launch 18 supercavitating NPs one-by-one toward the liquid interface from a distance (~4 nm) away from the interface (**Figure 4d).** We find that each of the NPs comes out of the liquid interface at a different angle and a different location (also see Supporting Movies, M5 and M6), similar to that observed in our experiments. The MD-simulated spread angles of these 18 NPs are plotted in the Supporting

Information, SI6.

**Conclusion**

In conclusion, we have demonstrated laser-driven photo-ejection of plasmonic NPs from liquid, which would not have been possible using conventional body forces. The laser with a wavelength at the SPR peak of the NPs can induce intense heating and form supercavitation. This process achieves the liquid-NP separation via a thermally induced phase change mechanism, and thus when the NPs are driven to the liquid free interface by the optical scattering force, they can move out of liquid without being stranded by the capillary force on the NP surface. Using finite element thermofluidic simulations, we prove that the observed NPs ejected out of liquid do not originate from any boiling or evaporation effect, confirming the supercavitation as the key. The NPs expelled out of liquid by the laser can be deposited on a substrate. By visualizing the deposited NPs, we observed a spreading angle larger than the divergence of the Gaussian laser beam, and this is attributed to the stochastic nature of nanobubble formation and the relative position of NP inside the nanobubble which results in fluctuations in the NP moving direction. This study reveals an interesting mechanism to separate NPs from a suspending liquid environment and could potentially lead to applications that utilize or need NP dry deposition and separation.

**Methods**

**Optical system to image the Au NPs photo-ejected out of liquid into air:** Figure 1a schematically shows the experimental setup used to probe the Au NPs ejected by laser from liquid into air. A femtosecond mode-locked monochromatic pulsed source laser (repetition rate of 80.7 MHz and pulse duration of 200 fs) from a Ti:sapphire crystal in an optical cavity (Spectra Physics, Tsunami) is directed to an Au NP suspension droplet. The center wavelength of the laser is 800.32 nm with a full-width-half-maximum of ~10.5 nm. The beam is focused by a 20× objective lens (Edmund Optics, numerical aperture = 0.42, focal length = 10 mm) onto the tip of the hemispherical droplet with a $1/e^2$ radius of 6 μm. The peak optical fluence at the focal point can be tuned continuously from 0.22 $mJ/cm^2$ to 22.8 $mJ/cm^2$ using a continuously variable metallic neutral density filter (NDC-25C-4M, Thorlabs). An optical shutter controls the on/off of the source laser. A high-speed digital camera (HX-7, NAC) with a 10× objective lens (Edmund Optics) is used to record the side view. A white LED (300 lm) illumination source is used for the bright-field imaging, and a coherent probe laser source with a wavelength of 632.8 nm (HeNe, 2 mW, Thorlabs) is used to image the NPs ejected from liquid in the dark field setting. The probe laser beam is directed to have an angle, ~90°, to the imaging axis of the high-speed camera (see **Figure 1a**) so that the probe laser cannot be seen by the camera directly, but the camera can capture the scattered probe laser light from the Au NPs ejected from liquid into air in the dark-field setting. We note that an optical filter is placed in front of the camera to filter out the source laser light (**Figure 1a**). The spatial distribution of NPs ejected into the air shown in **Figure 4c** is obtained by overlapping each frame of the captured video (similar to Supporting Movie M1) in 10 seconds with a customized image processing software in MATLAB.

**Substrate preparation and Au NP suspension:** Both the top and bottom substrates in **Figure 1a** are made of quartz. Before attaching the Au NP suspension droplet, the substrates were sequentially cleaned with acetone, isopropyl alcohol, ethanol and deionized water in an ultrasonic bath (PH30 Digital Ultrasonic Cleaner, Elma) and dried in a vacuum chamber (BACOENG). The Au NP (AuroShell, Nanospectra Biosciences, Inc.) used in this work consists of a silica core and an Au shell. The solvent for the Au NP suspension is deionized water produced by the Barnstead NanoPure Diamond system with a purity of 18 MOhm. The Au NPs have a near-infrared SPR wavelength at 780 ~ 800 nm. This near-infrared resonance wavelength coincides with the wavelength of the source laser we used in this work (~800 nm), that can induce enhanced plasmonic resonance to intensely heat up the NPs to excite the supercavitating nanobubbles.

**MD simulations:** To use a molecular model system to test the hypothesis that supercavitation can facilitate NPs moving out of liquid, we performed MD simulations for an Au NP immersed in liquid argon using LAMMPS (Large-scale Atomic/Molecular Massively Parallel Simulator).[60] Two contrasting cases respectively for a non-heated (90 K) NP and an intensely heated (1000 K) NP are simulated. Firstly, the system consisting of an Au NP immersed in liquid argon is created, energy-minimized and equilibrated in a canonical ensemble (NVT) at 90 K for 5 ns with periodic boundary conditions applied in all directions. Then, the simulation box size is increased in the z-direction to create a free space for the vapor phase. The whole structure is optimized in an isothermal-isobaric ensemble (NPT) at 1 bar and 90 K for another 5 ns to reach thermal equilibrium for the liquid-vapor two-phase system. For the non-heated case, the NP is given an initial velocity so it moves towards the liquid surface, and its movement is monitored. For the intensely heated case, after the system is fully relaxed at 90 K, we heat the NP to 1000 K via velocity rescaling with its center of mass fixed to achieve supercavitation. Then, the NP is given an initial velocity toward the liquid surface. We then monitor the resulting dynamical behavior to study whether the NP can move across the air/liquid interface. In all simulations, a solid slab at the bottom of the simulation cell is included and maintained at 90 K using a Langevin thermostat to dissipate heat so that the overall temperature of the whole system does not rise continuously. For the liquid, the Lennard-Jones (L-J) argon model,[61,62] $E_L(r) = 4\varepsilon\left[\left(\frac{\sigma}{r}\right)^{12} - \left(\frac{\sigma}{r}\right)^6\right]$, is used with $\sigma = 3.405$ Å and $\varepsilon = 0.24$ kcal/mol. For the NP, which has a radius of 1 nm, the Morse potential for Au[63] is used: $E_{NP}(r) = D_0\left[e^{-2\alpha(r-r_0)} - 2e^{-\alpha(r-r_0)}\right]$, where $D_0 = 10.954 \frac{\text{kcal}}{\text{mol}}$ is the bond-dissociation energy, $r_0 = 3.042$ Å is the equilibrium bond length, and $\alpha = 1.583$ Å is the parameter characteristic of the atom. The NP interacts with the liquid atoms also via the L-J potential with parameters $\sigma_{NP-L} = 3.405$ Å and $\varepsilon_{NP-L} = 0.46$ kcal/mol to model the hydrophilic surface. All interactions are truncated at $r_c = 12.5$ Å. It is worth mentioning that the distance between the NP with the slab is much greater than the force cutoff distance, so there is no influence from the slab on the dynamics of the NP. A time step of 5 fs is used for all MD simulations.

**Deposition of Au NPs photo-ejected from liquid:** A glass substrate is used to collect the Au NPs ejected from liquid under the droplet (**Figure 1a**). The distance between the bottom substrate and droplet is ~0.5 mm in point deposition (**Figures 3c-e**) and 200 μm in the line deposition (**Figure 4b**). Each deposition process lasts for 2 mins continuously. During line deposition, the bottom substrate is moved by a translation stage. To image the deposited Au NPs, a dark-field optical microscope (Olympus B-51) and a back-scattered SEM (Magelllan 400, FEI Company), with working voltage 10 kV, working current 0.2 ~ 0.4 nA and working distance 4.3 mm, are used. The sample is coated with a 2-nm Ir layer before SEM imaging and energy-dispersive X-ray (EDX) spectrum measurements.

**Acknowledgements**

This work is supported by National Science Foundation (1706039) and the Center for the Advancement of Science in Space (GA-2018-268). This work is supported by the National Research Foundation of Korea (NRF) grant funded by the Korea government (MSIT) (No. NRF-2021R1C1C1006251). We also appreciate the partial supports from Notre Dame Integrated Imaging Facility (NDIIF).

**Author contributions**

Q. Zhang and D. Huang contribute equally to this work. Q. Zhang, S. Moon, E. Lee and T. Luo designed the experiments, and Q. Zhang set up and performed the experiments. D. Huang, Q. Zhang, E. Lee and T. Luo designed the simulations. D. Huang and Q. Zhang performed the simulations. D. Huang and Q. Zhang wrote the manuscript, J. Schiffbauer, E. Lee and T. Luo revised it.

**Competing interests**

The authors declare no conflict of interest.